\documentclass{desyproc}

\newcommand{\be}{\begin{equation}}
\newcommand{\ee}{\end{equation}}

\newcommand{\ba}{\begin{eqnarray}}
\newcommand{\ea}{\end{eqnarray}}

\begin{document}


\title{Photon propagation in a cold axion condensate}

\author{{\slshape Dom\`enec Espriu and Albert Renau}\\[1ex]
Departament d'Estructura i Constituents de la Mat\`eria and Institut de Ci\`encies del Cosmos (ICCUB),
Universitat de Barcelona, Mart\'\i ~i Franqu\`es 1, 08028 Barcelona, Catalonia, Spain\\
}



\maketitle

\vspace{-7cm}\begin{flushright}{ICCUB-13-228}
\end{flushright}\vspace{6cm}

\begin{abstract}
We discuss some striking properties of photons propagating in a cold axion condensate oscillating coherently in
time with a frequency $1/m_a$. Three effects are discussed in this contribution: (a) due to the time dependence 
of the background, photons moving in the cold axion background have no definite energies and some momenta are
not accessible to them. (b) we investigate the combined influence of a magnetic field and the cold axion background
and propose a possible interferometric experiment to detect the latter. (c) if the axion
condensate has a space dependence, the photon refraction index is modified in the medium, possibly leading
to total reflection at the interface with the ordinary vacuum.

\end{abstract}

\section{Introduction}
Cold relic axions resulting from vacuum misalignment in the early universe are good candidates for 
dark matter\cite{sik,raff}. A coherent spatially constant axion field (it may be a genuine 
Peccei-Quinn axion\cite{PQ} or a similar field) acquires a mass $m_a$ once 
instanton effects set in and oscillates  
\begin{equation}\label{cos} a(t)= a_0 \cos m_a t.
\end{equation}
The energy density stored in these oscillations $\rho\simeq a_0^2 m_a^2$ contributes to the 
energy-matter budget of the universe. 
This constitutes the cold axion background, or CAB for short. 

As it is well known, the coupling of axions to photons is universal
\begin{equation}
\mathcal L_{a\gamma\gamma}=g_{a\gamma\gamma}\frac{\alpha}{2\pi}\frac a{f_a}F_{\mu\nu}\tilde F^{\mu\nu}.
\end{equation}
The only arbitrariness lies in the coefficient $g_{a\gamma\gamma}$ but most models
\cite{models} give $g_{a\gamma\gamma}\simeq 1$. Using \eqref{cos} 
\begin{equation} 
\mathcal L_{a\gamma\gamma}= -  g_{a\gamma\gamma} \frac{\alpha}{\pi} \frac{a_0}{f_a}
\cos (m_a t)\, \epsilon^{ijk} A_i F_{jk}.
\end{equation}
Cosmology considerations place the axion mass in a range $10^{-2} - 10^{-6}$ eV\cite{raff}. 
For Peccei-Quinn axions an 
approximate relation of the form $f_a m_a \simeq f_\pi m_\pi$ holds, thus forcing the axion decay constant
to be at least $f_a > 10^9$ GeV. From astrophysics the limit $f_a > 10^{7}$ GeV seems now well established.
This makes the axion extremely long-lived and very weakly coupled.

In addition to the axion-photon coupling, as indicated above, axions can also couple to 
matter in specific models. However, the smallness of the coupling makes their detection extremely challenging. 
Nevertheless, some of the best bounds on axion masses and couplings do come from the study of abnormal
cooling in white dwarfs due to axion emission\cite{wd}. 

In fact there are two separate questions we have to address. The first one is whether axions exist at all. This
is what experiments such as CAST, AMDX, AMDXII, IAXO or  ALPS try to address. If its mass turns out to be
in the relevant range for cosmology we would indeed have a strong hint that axions may provide the elusive
dark matter. 

However, we would eventually like to verify or falsify the CAB hypothesis and determine to what extent cold axions
contribute to the dark matter budget. Finding the axion particle with the appropriate properties
is not enough. Unfortunately a direct experimental confirmation of the CAB is extremely difficult 
and it could possibly be accomplished only via some collective effect on the propagation of particles. 
Looking for the effect of the CAB on photon propagation 
is the most natural possibility.

In a previous Patras meeting we reported on three effects that the presence of a CAB induces on photon propagation\cite{patras}.
The CAB modifies the photon dispersion relation introducing a Lorentz non-invariant term, which makes
Bremsstrahlung from cosmic rays possible. The amount of energy radiated by this process was computed 
and found to be non-negligible\cite{ours}, but normal synchrotron radiation background is a tough enemy for its detection.  
The second effect discussed was the presence of 
an additional rotation in the polarization plane of light 
(on top of the familiar one\cite{raffelt}, see \cite{last} for a detailed discussion). 
The third effect discussed is that some photon wave-lengths 
are actually forbidden in a universe filled with cold axions oscillating coherently\cite{last}.
In this talk we will continue our discussion on the last two points, and also speculate very briefly on possible 
consequences of having a CAB with some spatial dependence.

In order to determine the properties of photons in a CAB we need to solve the equation of motion in momentum space.
For the time being let us ignore the magnetic field.
\begin{equation} \left[g^{\,\lambda\nu}\left(k^2-m^2_\gamma\right) +
i \,\varepsilon^{\,\lambda\nu\alpha\beta}\,\eta_\alpha\,k_\beta\right]
\tilde A_\lambda(k)=0.
\label{movement}
\end{equation}
where $\eta_\alpha\sim  \partial_\alpha a= \delta_{\alpha 0} \dot a$. We shall approximate the 
sinusoidal variation with of the axion background by a piecewise linear function (see figure). 
The astrophysical and observational bounds 
on  $\vert \eta_0\vert =\pm 2g_{a\gamma\gamma}\frac\alpha\pi\frac{a_0m_a}{f_a}$ range from  $ \vert \eta_0\vert < 10^{-24}$ to
$ \eta_0 < 10^{-20}$ eV.  The quantity $\eta_0$ changes alternatively
from positive to negative with a period $2\pi/m_a$. Within each period of oscillation 
two complex and space-like chiral polarization vectors 
$\varepsilon^{\mu}_{\pm}(k)$ can be defined (see \cite{sasha}). 
The two polarization vectors
are solutions of the vector field equations if and only if
\begin{equation} k^{\mu}_{\pm}=(\omega_{\pm} , {\vec k})\qquad
\omega_{\pm}=\displaystyle
\sqrt{{\vec k}^2+m_{\gamma}^{2}\pm\eta_0 k},\qquad  k=\vert \vec k\vert.
\label{reldis}\end{equation}

\section{Forbidden wavelengths}
Using this approximation of replacing the CAB oscillation by a piecewise-linear function, as shown in the figure,
we can solve exactly for the propagating modes.
\begin{figure}[h]
\center
\includegraphics[scale=0.7]{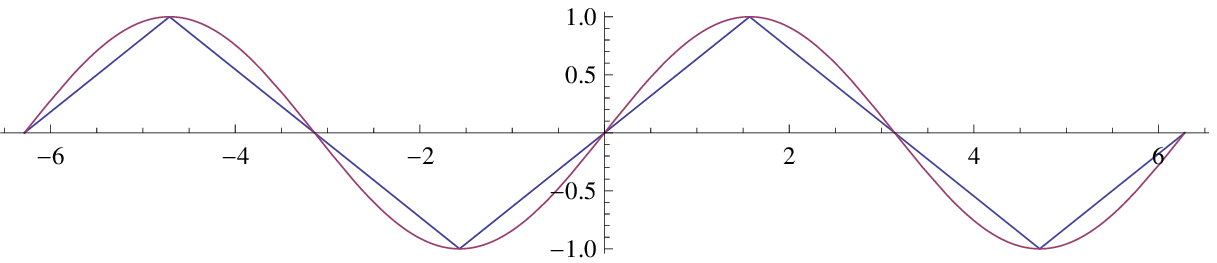}
\label{tsin}\end{figure}
The equation for $\hat A_\nu(t,\vec k)$ is
\be\label{eq}
\left[g^{\mu\nu}(\partial_t^2+\vec k^2)-i\epsilon^{\mu\nu\alpha\beta}\eta_\alpha k_\beta\right]\hat A_\nu(t,\vec k)=0.
\qquad
\hat A_\nu(t,\vec k)=\sum_{\lambda=+,-}f_\lambda(t)\varepsilon_\nu(\vec k,\lambda).
\end{equation} 
We write $f(t)=e^{-i\omega t}g(t)$ and demand that $g(t)$ have the same periodicity as $\eta(t)=\eta_0\sin(m_at)$. This
requires the fulfillment of the condition
\begin{equation} 
\cos(2\omega T)=\cos(\alpha T)\cos(\beta T)-\frac{\alpha^2+\beta^2}{2\alpha\beta}\sin(\alpha T)\sin(\beta T),
\quad T= \frac{\pi}{m_a},
\end{equation} 
where $\alpha,\beta$ coincide with the two frequencies $\omega_{\pm}$.

A close inspection of the solutions to the previous equation reveals the existence of momentum gaps:
some values of $k$ admit no solution for $\omega$. The phenomenon is visible for large values
of the dimensionless ratio $\eta_0/m_a$.
\begin{figure}[h]
\center
\includegraphics[scale=0.3]{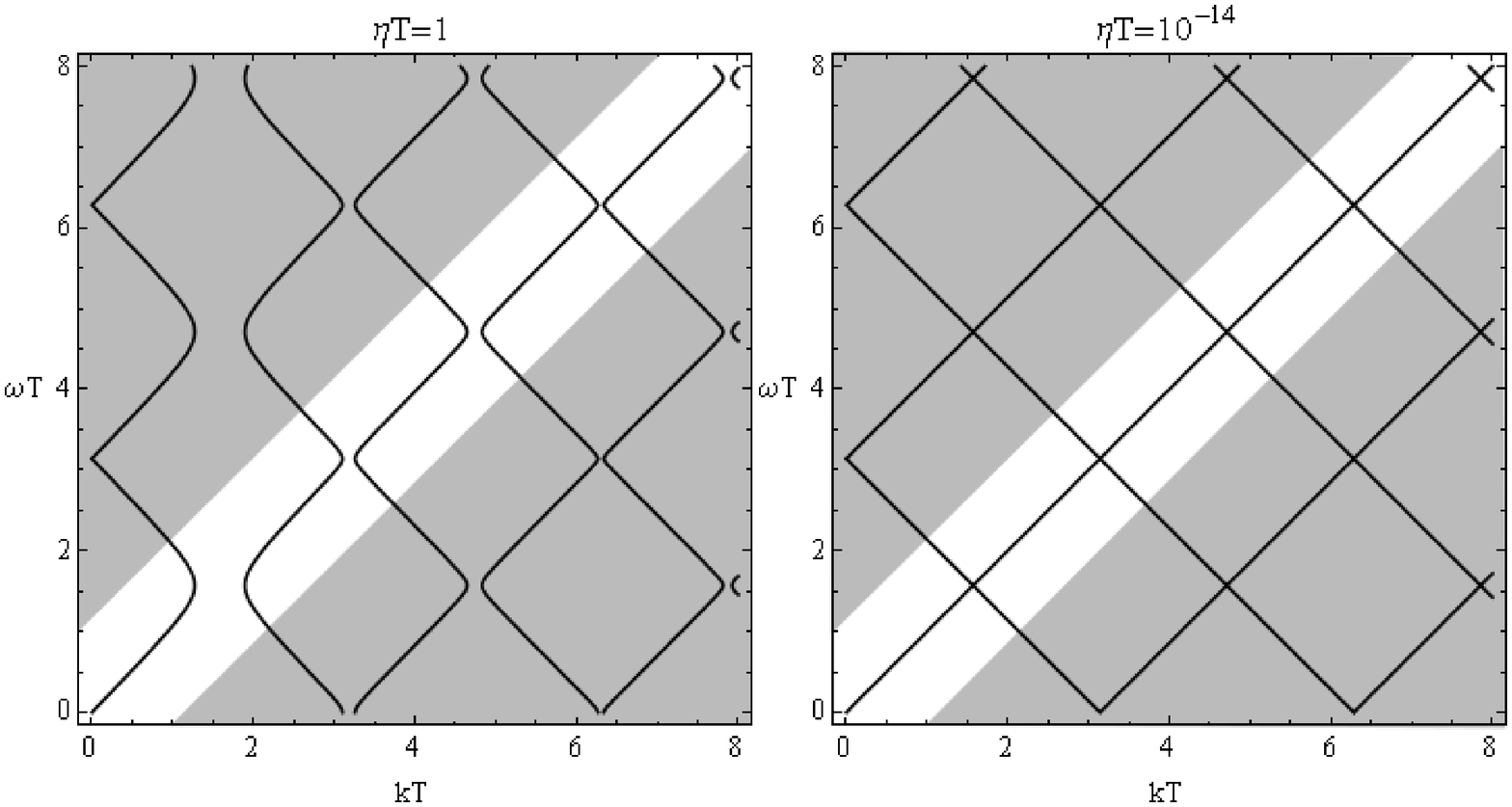}
\end{figure}
In a universe filled with cold axions oscillating with a period
$2\pi/m_a$ some wavelengths are forbidden by a mechanism that is similar to the 
one preventing some energies from existing 
in a semiconductor.  See \cite{last} for more details. 
Unfortunately, for realistic values of $\eta_0$ the gaps are probably far too small to be seen. Their width is actually
proportional to $\eta_0^2/m_a$

\section{Adding a magnetic field}
If $\eta_0=0$ the theoretical technology is well known. It amounts to resumming the Feynman diagrams shown below.
\begin{figure}[h]
\center
\includegraphics[scale=0.3]{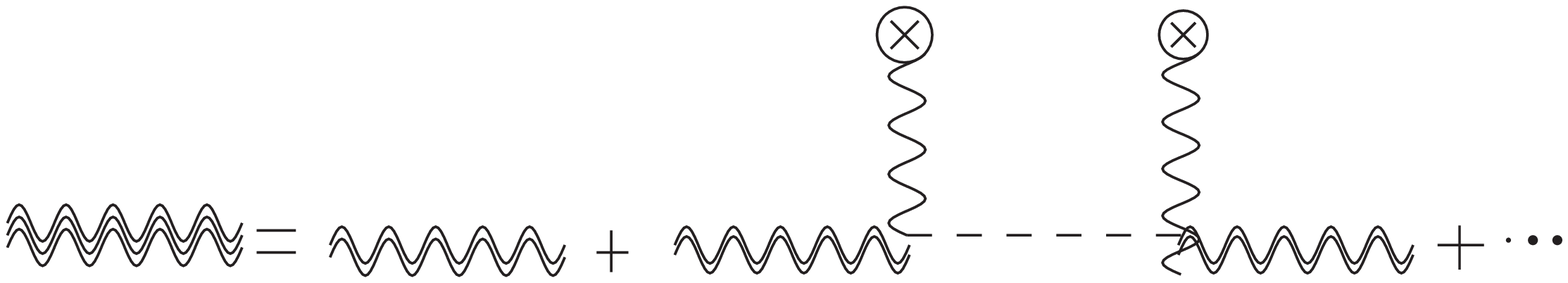}
\end{figure}
Interaction with the cold axion background implies that we need to take into account also
the diagrams that include interaction with the CAB
\begin{figure}[h]
\center
\includegraphics[scale=0.3]{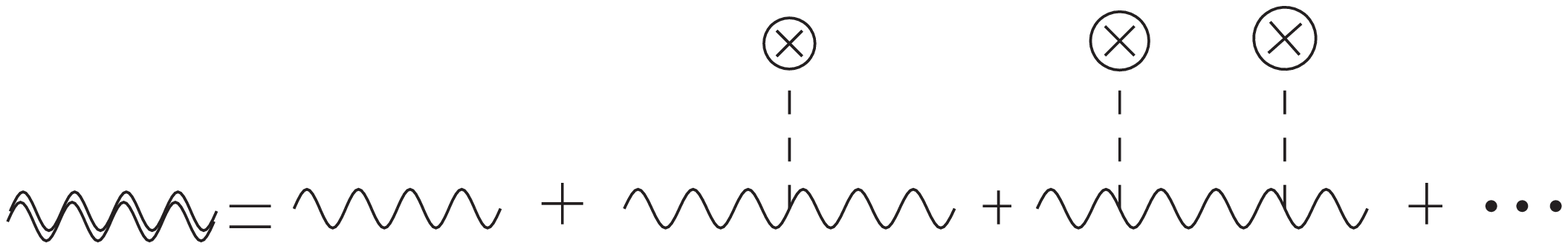}
\end{figure}
This can be done and the resulting resummed propagator can fe found in \cite{last}, where we
examined the evolution of the polarization plane under the joint influence of the magnetic field and
the CAB. 

The relevant parameters are $b= 2 g_{a\gamma\gamma} \frac{\alpha}{\pi}\frac{B}{f_a}$ and $\eta_0$, where
$B$ is the magnetic field. Taking 
as a reference value (just to set an approximate range of variation for $b$)  
$f_a> 10^7 $ GeV (this would correspond for PQ axions to $m_a \simeq 1$ eV) and $ B= 10\ {\rm T}$ 
then  $b \le 10^{-15} \,{\rm eV}$ and
$\eta_0 \le 10^{-20} \,{\rm eV}$. However, we shall consider $m_a$ and $f_a$ (and obviously $b$) to be free
parameters.

The relevant evolution equations are
\be
\left(
\begin{array}{ccc}
 \partial_t^2+k^2+m_a^2 &   -ib\partial_t    & 0        \\
 -ib\partial_t          & \partial_t^2+k^2 & \eta_0 k                    \\
 0      & \eta_0 k                & \partial_t^2+k^2
\end{array}
\right)
\left(
\begin{array}{c}
 \hat a \\ i\hat A_1 \\ \hat A_2
\end{array}
\right)
=
\left(
\begin{array}{c}
 0 \\ 0 \\ 0
\end{array}
\right) 
\ee
We try a solution of the form: $(\hat a,i\hat A_1, \hat A_2)=e^{-i\omega t}(x,X_1,X_2)$:
\be
\left(
\begin{array}{ccc}
 -\omega^2+k^2+m_a^2 & \omega b         & 0         \\
 \omega b          & -\omega^2+k^2 & \eta_0 k                    \\
 0      & \eta_0 k                & -\omega^2+k^2
\end{array}
\right)
\left(
\begin{array}{c}
x\\ X_1 \\ X_2 
\end{array}
\right)
=
\left(
\begin{array}{c}
 0 \\ 0 \\ 0
\end{array}
\right)  
\ee
Within each period of oscillation the proper frequencies are in the limit $\eta_0\ll b\ll\{m_a,k\}$
\ba 
\omega_a^2&\approx& k^2+m_a^2+b^2+\frac{b^2 k^2}{m_a^2}+\frac{b^2 \eta_0^2 k^4}{m_a^6}+\frac{b^2 \eta_0^2 k^2}{m_a^4}\\ 
\omega_1^2&\approx& k^2-\frac{b^2 \eta_0^2 k^4}{m_a^6}-\frac{b^2 \eta_0^2 k^2}{m_a^4}-\frac{b^2 k^2}{m_a^2}-\frac{\eta_0^2 m_a^2}{b^2}\\
\omega_2^2&\approx& k^2+\frac{\eta_0^2 m_a^2}{b^2}
\ea
To be compared with the proper modes without magnetic field
\be
\omega^2_\pm = k^2 \pm \eta_0 k .
\ee
Note that one should not attempt to take  the $b\to 0$ limit above because we assume $\eta_0 \ll b$.
The natural basis is now parallel and perpendicular (to the magnetic field) rather than left or right polarization. 
Notice that the proper frequencies in this natural basis and in this limit contain only even 
powers of  $\eta_0$ and hence do not change
when $\eta_0$ changes sign. Unlike the $b=0$ case both modes are in an eigenstate of energy. We do not expect momentum gaps.

Numerically for $b= 10^{-15}$ eV (a magnetic field of $10$ T for $f_a=10^{7}$ GeV), 
$\eta_0= 10^{-24}$ eV and $k= 1$ keV, the relevant terms are
 \ba \nonumber
\omega_a^2&\approx& k^2+m_a^2+\frac{b^2 k^2}{m_a^2}\\ 
\omega_1^2&\approx& k^2-\frac{b^2 k^2}{m_a^2}-\frac{\eta_0^2 m_a^2}{b^2}\\ 
\omega_2^2&\approx& k^2+\frac{\eta_0^2 m_a^2}{b^2} .
\ea
The splitting between the two polarizations goes as 
\be\label{xurro}
\omega_2^2-\omega_1^2 \approx \frac{b^2k^2}{m_a^2} + 2 \frac{\eta_0^2 m_a^2}{b^2} .
\ee
The relevance of the CAB is somehow enhanced by the magnetic field and dominates for
\be
b < m_a \sqrt\frac{\eta_0}{k} \simeq 10^{-14} m_a.
\ee 
The difference (\ref{xurro}) is very small; typically it
could be as ``large'' as $10^{-9}$eV$^2$. Small as this number is, it should be remembered that
Michelson-Morley type experiments can detect differences in frequencies with 17 significant figures\cite{MM}.
If so, by tuning the magnetic field one should be able to switch in and out the effect of the CAB and
perhaps ``seeing'' it.

There is also a change in the plane of polarization with an angle
 \be
\sim \frac{\eta_0 m_a^2}{b^2k}
\ee
(for ``large'' axion masses, the actual result is more complicated).
This angle could be be as large as $10^{-3}$. What this means is that the plane of polarization changes 
from period to period. Of course the change is not instantaneous and a more complete 
description of the evolution can be found with the help of the photon correlator 
derived in \cite{last}. The evolution of the plane of polarization with respect to the $b=0$ case 
has a new component due to the CAB.

The rotation survives even without magnetic field and the effect is independent of the frequency.
Note that the previous result holds only for table-top experiments when the photon can approximately be considered
an eigenstate of energy, i.e. when the time-of-flight of the photon, $\vert x \vert $,  is smaller than $2\pi /m_a$.
This means $\vert x \vert < 2\pi /m_a$ and the method could perhaps be useful for very small axion masses.

Due to the smallness of the numbers involved  it is somewhat dangerous to rely on approximate formulae - many 
scales play in the game. In a forthcoming paper\cite{next} a detailed numerical analysis will be performed.

\section{Crossing the boundary}
Let us finish with a few words on the third topic that was discussed in our presentation. This is work done
in collaboration with A. Andrianov and S. Kolevatov and it will be described very succintly here. We refer
the readers to \cite{AK} for more details.

We want to explore different possible axion backgrounds (other than the cold background
oscillating in time with  period $\sim 1/m_a$). Consider the term in the lagrangian
\be
-\,{\textstyle\frac14}\,F^{\mu\nu}(x)\widetilde F_{\mu\nu}(x)\,\zeta_\lambda x^{\lambda}\,\theta(-\,\zeta\cdot x)
 \leftrightarrow {\textstyle\frac12}\,\zeta_\mu A_\nu(x)\widetilde F^{\mu\nu}(x)\,\theta(-\,\zeta\cdot x),
\ee
This associates a space-like boundary with a space-like vector 
\be
\zeta_\mu = \zeta \times (0,\vec a)\quad |\vec a|= 1 
\ee
The Lorentz invariance violating  vector 
has been renamed from $\eta_\mu$ to $\zeta_\mu$ to avoid confusion with the CAB case.

Although not totally realistic for astrophysical purposes, this would correspond to a region where a linearly 
decreasing axion density meets the vacuum. For convenience the ``wall'' is placed in the $\hat X$ direction at $x=0$.
Matching on the boundary  $\zeta\cdot x = 0$ leads to 
\be
\delta(\zeta\cdot x)\left[ \,A^{\mu}_{\rm vacuum}(x)-A^{\mu}_{\zeta}(x)\,\right] = 0.
\ee
Different polarizations have different dispersion relations in the axion phase
\begin{eqnarray}
\left\lbrace
\begin{array}{cc}
k_{1L}=k_{10}=\sqrt{\omega^2-m_\gamma^2-k_\bot^2}\\
k_{1+}=\sqrt{\omega^2-m_\gamma^2-k_\bot^2+\zeta \sqrt{\omega^2-k_\bot^2}}\\
k_{1-}=\sqrt{\omega^2-m_\gamma^2-k_\bot^2-\zeta \sqrt{\omega^2-k_\bot^2}}\\
\end{array}\right.
\end{eqnarray}
that have to be matched to the usual dispersion law in the normal phase ($m_\gamma$ is the photon
mass and need not be zero, due to plasma effects)
\be
k_1 = \sqrt{\omega^2-m_\gamma^2-k_\bot^2}. 
\ee
This actually leads to a non-trivial reflection coefficient (see \cite{AK} for details).
As a function of the invariant mass of the photon $M^2$ 
\be
\kappa_{ref} (M^2)= 
\frac{\vert\sqrt{\frac{(M^2-m_\gamma^2)^2}{\zeta^2}-M^2}-\sqrt{\frac{(M^2-m_\gamma^2)^2}{\zeta^2}-m_\gamma^2}\vert}
{\vert\sqrt{\frac{(M^2-m_\gamma^2)^2}{\zeta^2}-M^2}+\sqrt{\frac{(M^2-m_\gamma^2)^2}{\zeta^2}-m_\gamma^2}\vert}
\ee
Again, the effect seems to depend crucially on the ratio of two very small numbers: $m_\gamma$ and $\eta_0$.

\section*{Acknowledgments}

It is a pleasure to thank the  organizers of Patras 2013 for an enjoyable conference and encouragement. 
We acknowledge financial support from projects FPA2010-20807, 2009SGR502 and Consolider CPAN.

\end{document}